\documentclass[preprint,aps,floatfix]{revtex4}

\usepackage{graphicx}
\usepackage{dcolumn}
\usepackage{bm}

\newcommand{\pdiffl}[2]{\frac{\partial #1}{\partial #2}}

\begin{document}


\title{Shock formation and the ideal shape of ramp compression waves}

\date{May 29, 2008 -- LLNL-JRNL-404275}

\author{Damian C. Swift}
\email{dswift@llnl.gov}
\affiliation{%
   CMELS-MSTD, Lawrence Livermore National Laboratory,
   7000 East Avenue, Livermore, California 94551, USA
}
\author{Richard G. Kraus}
\affiliation{%
   Department of Physics, Cavendish Laboratory, University of Cambridge,
   JJ~Thomson Avenue, Cambridge CB3~0HE, UK
}
\author{Eric Loomis}
\affiliation{%
   Group P-24, Los Alamos National Laboratory,
   Mail Stop E526, Los Alamos, NM 87545, USA
}
\author{Damien G. Hicks}
\affiliation{%
   PS-V, Lawrence Livermore National Laboratory,
   7000 East Avenue, Livermore, California 94551, USA
}
\author{James M. McNaney}
\affiliation{%
   CMELS-MSTD, Lawrence~Livermore National Laboratory,
   7000 East Avenue, Livermore, California 94551, USA
}
\author{Randall P. Johnson}
\affiliation{%
   Group P-24, Los~Alamos National Laboratory,
   Mail Stop E526, Los Alamos, NM 87545, USA
}

\begin{abstract}
We derive expressions for shock formation based on the local curvature of the
flow characteristics during dynamic compression.
Given a specific ramp adiabat, calculated for instance from the equation of
state for a substance, the ideal nonlinear shape for an applied ramp loading
history can be determined.
We discuss the region affected by lateral release, which can be presented
in compact form for the ideal loading history.
Example calculations are given for representative metals and plastic ablators.
Continuum dynamics (hydrocode) simulations were in good agreement
with the algebraic forms.
Example applications are presented for several classes
of laser-loading experiment,
identifying conditions where shocks are desired but not formed, and where long
duration ramps are desired.
\end{abstract}


\maketitle

\section{Introduction}
Lasers are used increasingly in the study of the response of matter under
extreme conditions, by inducing dynamic loading by ablation.
The canonical classes of dynamic loading experiment are 
shocks \cite{shockref} and ramps \cite{rampref}.
Laser ablation can be used to induce either type of loading by altering the
irradiance history of the laser pulse \cite{Swift_lice_05}.
To a crude approximation, the ablation pressure is proportional to the
irradiance \cite{Swift_elements_04}, so a square laser pulse induces a shock 
and a ramped pulse, a ramp.
A shock wave propagating through matter has an inherent rise time, related
to the nature of the dissipative processes, such as viscosity or scattering, 
causing the associated increase of entropy.
For a simply-behaved material (neglecting or simplifying 
time-dependent responses such as plastic flow and phase changes),
a shock propagates unchanged if the drive supports it for long enough,
whereas a ramp of a given rise time progressively steepens as it propagates.
Eventually, the rise time reaches the inherent rise time of a shock,
and the ramp becomes a shock.
Laser pulses generally have a finite rise time even when a shock
is intended, so some part of the target material is subjected to a ramp until
it steepens to form a shock.

Here we consider the formation of a shock from a ramp, with application to
several situations in laser ablation experiments.
We consider several different classes of laser-shock experiment, 
discussed later in more detail,
but broadly depending on the pulse energy of the laser.
Experimental techniques have been developed furthest for high energy lasers,
and here we are interested in understanding how much of the sample is not
shocked, for comparisons with microscopic analysis of recovered samples
\cite{Peralta05}.
A current interest is the use of lasers of lower energy that can be transported
to other facilities to induce loading which is then probed by other techniques
such as synchrotron radiation \cite{McNaney08};
here we want to determine whether any given laser system is capable of producing
a shock in any useful part of the sample.
Finally, we are interested in the optimization of ramp loading experiments to
allow ramp loading to occur over the maximum possible distance
before a shock forms.

In principle, all of these situations can be investigated using 
spatially-resolved continuum dynamics simulations: hydrocode calculations.
However, the numerical time-integration algorithms in these simulations 
are almost universally unstable when the solution contains a shock wave,
which is a perfect discontinuity in the continuum approximation.
An artificial viscosity is used to smear a shock over several spatial zones
\cite{Benson92}.
The use of shock smearing makes it difficult to study the formation of
a shock from a ramp, because it removes the clear distinction between 
the different types of wave, and the shock formation process may depend on
the specific form of artificial viscosity chosen.
Here we analyze the steepening of a ramp and the formation of a shock
in terms of the characteristics of the flow, which does not require a numerical
discretization of space (or time), and allows shock formation to be
identified uniquely in the continuum approximation.

As discussed below, the steepening of a ramp compression wave is closely
connected with the spreading of a release wave, and can be investigated using
the same relations.

\section{Steepening of a ramp wave}
\begin{figure}
\begin{center}\includegraphics[scale=0.72]{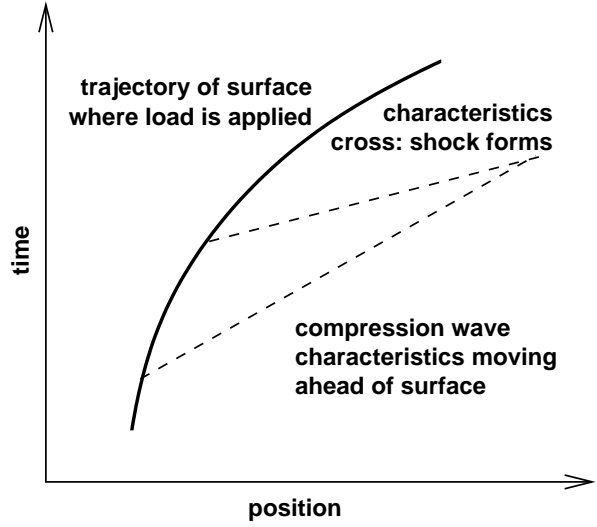}\end{center}
\caption{Schematic of the crossing of characteristics in a ramp wave,
   signaling the formation of shock wave.}
\label{fig:steepschem}
\end{figure}

A ramp wave evolves as the region at a given pressure engulfs more material,
the incremental compression wave traveling at the instantaneous sound speed
$c$.
In a material whose equation of state (EOS) is simple, $c$
increases monotonically with pressure $p$.
Thus the ramp wave steepens as it propagates.
The steepening can be understood in terms of characteristics of
the continuum equations, which for a material described by a scalar EOS
comprise in one dimension the material (or particle) flow velocity $u$ and
sound waves propagating forward and backward with respect to the flow,
$u\pm c$.
While the flow remains ramp-like, the characteristics continue
as straight lines in position-time space.
If a pair of forward- or backward-propagating characteristics crosses,
a shock forms.
In general, the shock does not initially encompass the full pressure range 
of the ramp or even its limits; a ramp may form an embedded shock over any
part of its range, and the shock may then spread upward and downward in
pressure.
Other parts of the ramp may form a shock independently before the 
first-forming shock engulfs them.
(Fig.~\ref{fig:steepschem}.)

We consider the formation of a shock by the crossing of characteristics 
in any part of a ramp compression.
We consider two derivations, Lagrangian and Eulerian (taken here to
mean respectively coordinate systems moving with the material or fixed
in space \cite{Benson92});
the alternative derivations are equivalent but lead to different
expressions for the steepening of a ramp that are more convenient in different
situations.

As a point of terminology, ramp compression is commonly referred to
as isentropic or quasi-isentropic.
If the material is represented by an inviscid, time-independent scalar EOS,
ramp compression follows an isentrope.
This is reasonable when dissipative processes such as
irreversible plastic work,
and time-dependence in, for instance, plasticity and phase changes,
can be neglected.
The analysis presented here is valid for more general material behavior,
and we refer to the thermodynamic trajectories as adiabats 
since no heat is exchanged with
the surroundings on the time scales of interest.

\subsection{Lagrangian derivation}
In a frame of reference moving with respect to an element of deforming
material,
the speed of a longitudinal sound wave at the local compression
(mass density $\rho$ and pressure $p$) is $c$.
As the compression increases in a ramp, $c$ changes.
The distance the ramp must propagate for a shock to form at $p$
is derived by considering the speed of successive characteristics
starting at different times: in incremental form $c(p)$ starting at time $t$,
and $c(p+\delta p)$ at $t+\delta t$,
where $\delta p/\delta t$ is the pressurization rate $\dot p(p)$.
The compression wave at higher pressure travels a shorter distance
through compressed material, which can be accounted for by considering
the intersection in a coordinate frame fixed with respect to the 
undeformed material (an alternative meaning of `Lagrangian'),
where the longitudinal sound speed is $c\rho/\rho_0$.
The distance $l_0$ for the shock to form in the part of the ramp wave at $p$,
expressed in terms of the {\it uncompressed} material, is given by
\begin{equation}
\tilde\rho\equiv\dot\rho l_0
   =\frac 1{\rho_0}\frac{\rho^2c^2}{c+\rho\left.\pdiffl c\rho\right|_s}
\end{equation}
where tilde quantities are time derivatives multiplied by the uncompressed mass
density $\rho_0$, i.e. scaled rates of change.
Given the longitudinal sound speed along the adiabat, $c(\rho)$, 
this relation can be used to
calculate a scale time $\tau=t/l_0$ for shock formation, 
as a function of $\rho$.
Then given the pressure along the adiabat, $p(\rho)$,
the scaled pressurization rate $\tilde p(p)$ can be calculated 
and hence $p(\tau)$.
For material described by a scalar EOS,
these relations can be expressed in terms of the bulk modulus $B$ 
and its derivative, since $c^2=B/\rho$.
Loading rates expressed in terms of scaled time are natural when time-dependent
processes can be neglected 
(such as the kinetics of phase changes and plastic flow),
as is often the case for applications in any given regime of loading rate,
as the continuum dynamics equations are self-similar.
This formulation is then convenient because it
captures the shock formation process compactly irrespective of the actual
loading rate, as a property derived from a given adiabat with no further
assumptions about or constraints on the loading history.

This result is similar to a previous derivation \cite{Davis05},
except that we consider the instantaneous curvature of each characteristic
with respect to pressure, rather than
their intersection after a formally finite compression.
The derivation presented here therefore
can be used for more general loading conditions,
such as ramp following an initial shock,
or reflected from an impedance mismatch.
A difference in convention is that we avoid
formulation in terms of the `Lagrangian sound speed'
($C\equiv\rho c/\rho_0$: the speed with respect to the uncompressed material),
as this not a helpful quantity in general loading scenarios or for
general material models (such as porous materials).
The sound speed is defined more naturally with respect to moving material in its
instantaneous state of compression and deformation.

For materials described by constitutive models (including EOS) of arbitrary
complexity, the adiabat can be calculated as a numerical tabulation 
$\{\rho,p,c\}$ \cite{Swift-genscalar-07}.
The finite difference version of the shock formation relation gives the
increment in scaled time between adjacent states $i$ and $i+1$ in the table:
\begin{equation}
\Delta\tau\equiv\frac{\Delta t}{l_0}
   = \rho_0\left(\frac 1{\rho_i c_i}-\frac 1{\rho_{i+1}c_{i+1}}\right).
\end{equation}

\subsection{Eulerian derivation}
The distance and time to form a shock from an increment of compression
around a pressure $p$ can be calculated similarly by the intersection of
characteristic in the laboratory frame.
The speed of a longitudinal sound wave in the laboratory frame is $u+c$,
where $u$ is the instantaneous velocity of the material in that part of the
ramp.
Now the intersection takes account that characteristics starting from a
given point in the material at different times move, so
$u$ is integrated to find the position $x(t)$.
The Eulerian derivation is less elegant for scaled quantities, 
but is expressed in terms of $u$ rather than $\rho$,
which is convenient for some applications,
such as when analyzing the properties of a loading history
predicted by some types of continuum dynamics simulation 
\cite{Swift-heflyer-07}.
Again considering intersection at a distance $l_0$ into the stationary material,
The rate of change $\dot\sigma$ of any state parameter $\sigma$
is expressed in terms of the distance $l_0$ in the laboratory frame 
for the characteristics around $\sigma$ to cross:
\begin{equation}
\dot\sigma
=\frac{(l_0-x)c(u+c)}{\left.\pdiffl u\sigma\right|_s+\left.\pdiffl c\sigma\right|_s}.
\end{equation}
Particularly useful state parameters are $\rho$ and $p$, as they can readily
be determined along an adiabat, as discussed above.
This relation allows the time-derivative to be obtained from the
rate of change of any state parameter along the adiabat.

For a tabulated adiabat $\{u,p,c\}$,
the time increment between adjacent states is
\begin{equation}
\Delta t
   =\frac{l_0-(x_i+x_{i+1})/2}{u_i+c_i}
      \frac{u_{i+1}+c_{i+1}-u_i-c_i}{u_{i+1}+c_{i+1}-(u_i+u_{i+1})/2},
\end{equation}
which yields a scaled time increment $\Delta\tau\equiv\Delta t/l_0$ 
by choosing $l_0=1$.

Calculations using the Lagrangian and Eulerian derivations give 
identical results,
as do calculations using the derivative and difference formulations
of the relations.

\subsection{Analytic solution for a perfect gas}
The ramp steepening relations can be expressed in analytic form for sufficiently
simple forms of EOS.
The perfect gas EOS, $p=(\gamma-1)\rho e$, gives isentropes satisfying
$p/\rho^\gamma$ constant.
The sound speed $c=\sqrt{\gamma p/\rho}$.
Thus the Lagrangian formulation gives
\begin{equation}
\tilde\rho_{\mbox{perfect gas}}=\frac{2\rho^2c}{(\gamma+1)\rho_0}.
\end{equation}
The Eulerian relation can be verified similarly by using the relation
$\partial u/\partial\rho=c/\rho$ along the isentrope.

\section{Ideal shape of ramp waves in selected materials}
If the evolution of a ramped loading history is to be used in an experimental
study of material properties, a common requirement is to design the ramp
so that the first shock forms as late as possible, 
i.e. allowing as thick a region of material as possible
to be subjected to a pure ramp as opposed to a shock
over any part of the compression range.
For a given overall rise time of the the ramp, the ideal shape is the one 
where the shock forms simultaneously over the whole pressure range,
i.e. the characteristics all cross at the same position and time.
Because of hydrodynamic scaling in situations with negligible time-dependence
in the response of the materials to loading, the ideal ramp shape is
self-similar with respect to time before the formation of the shock.
In other words, `running time backwards' from the instant at which the
shock forms, the ramp wave progressively broadens, 
or its history at any Lagrangian point 
(a distance $l_0$ from the shock formation position, in unshocked material)
gives the ideal loading history
to apply in order to form a shock simultaneously after compressing a thickness
$l_0$ of material.

The same analysis can be applied to the spreading of a release wave when 
an applied pressure is abruptly relieved, as at the end of a laser drive
pulse or when an impact-induced shock reaches an interface with a material
of lower impedance (such as a free surface).
In this case, the release adiabat from the high pressure state is used
rather than the adiabat starting at the ambient state.

These calculations do require the constitutive properties of the material to be
known or estimated, so estimates are needed when designing experiments to
investigate unknown properties.
The analysis described above is, however, valid for general material models
including strength, as long as the adiabat can be calculated
\cite{Swift-genscalar-07}.
The examples shown below are however for materials represented by
a scalar EOS, where the ramp adiabat is an isentrope.

\begin{figure}
\begin{center}\includegraphics[scale=0.72]{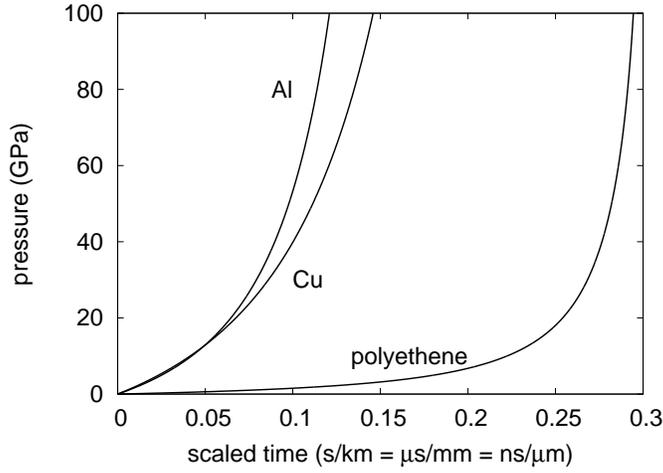}\end{center}
\caption{Ideal scaled loading histories for Al, Cu, and polyethene.}
\label{fig:tp}
\end{figure}

\begin{figure}
\begin{center}\includegraphics[scale=0.72]{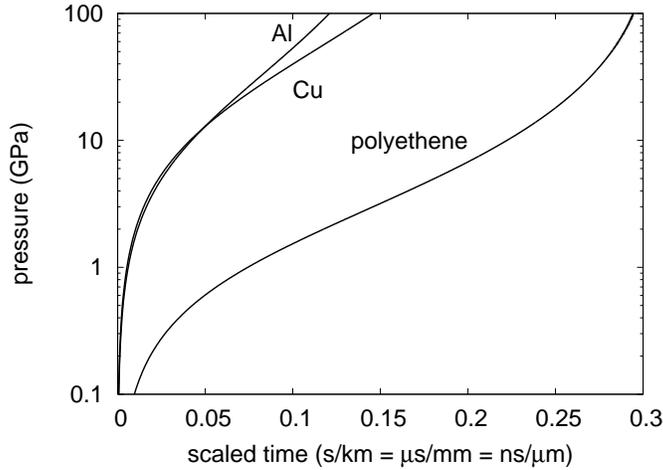}\end{center}
\caption{Ideal scaled loading histories for Al, Cu, and polyethene
   (log space).}
\label{fig:tlogp}
\end{figure}

Material properties were taken from a compendium of parameters for 
analytic models, fitted to experimental data
\cite{Steinberg96}.
The EOS used a polynomial fit to shock speed data, and a density-dependent
Gr\"uneisen parameter for off-Hugoniot states.
This model is unphysical at high ramp compressions when states are far 
from the reference Hugoniot curve.
Calculations were performed for Al and Cu, as prototype metals of very different
ambient mass density, and also for polyethene, which we have shown
previously \cite{Swift-chablator-07} is a reasonable prototype 
ablator material as used in some types of laser-driven ramp experiment.
The results are plotted as pressure as a function of scale time, i.e. $p(\tau)$
(Figs~\ref{fig:tp} and \ref{fig:tlogp}).
To interpret these graphs as real time, choose a thickness for the shock
to form ($l_0$) and multiply $\tau$ by $l_0$.
Thus, to design an experiment loading Cu to 80\,GPa (scale time approximately
0.13\,ns/$\mu$m) where the distance to form a shock should be at least
100\,$\mu$m say, time $t=100\tau$\,$\mu$m and the drive should take
at least 13\,ns to reach 80\,GPa.

\begin{figure}
\begin{center}\includegraphics[scale=0.72]{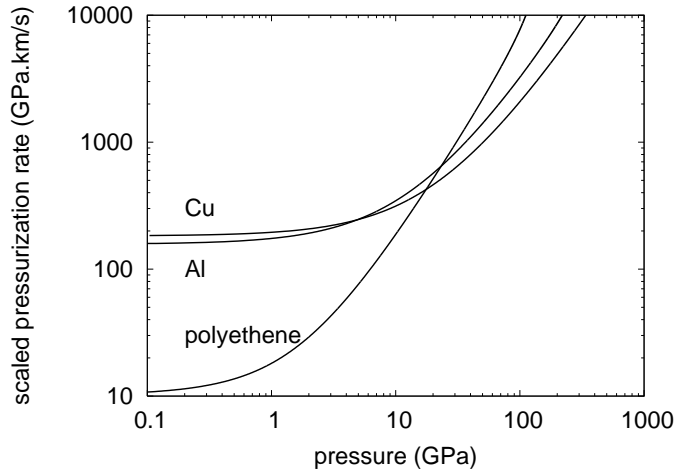}\end{center}
\caption{Ideal scaled loading rate as a function of instantaneous pressure,
   for Al, Cu, and polyethene.}
\label{fig:pptilde}
\end{figure}

\begin{figure}
\begin{center}\includegraphics[scale=0.72]{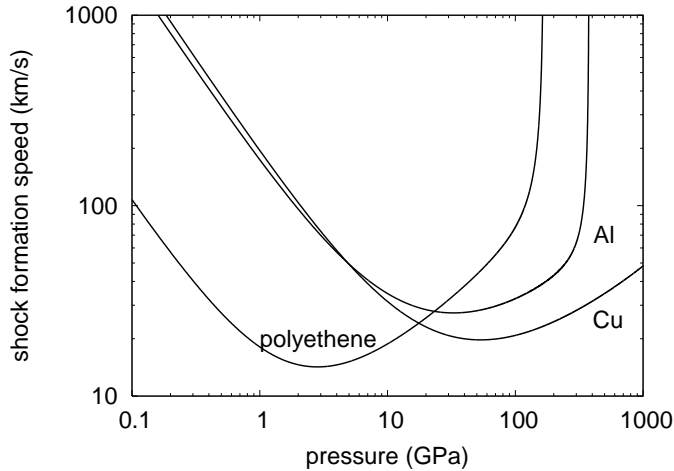}\end{center}
\caption{Ideal scaled loading rate as a function of instantaneous pressure,
   for Al, Cu, and polyethene.
   The abscissa is that of the previous graph, divided by the ordinate.
   This gives a convenient measure of the systematic trend of the
   overall pulse duration for ideal ramp loading to a given pressure,
   but not an accurate estimate because the ideal shape is non-linear
   with monotonically increasing rate.}
\label{fig:ptilde}
\end{figure}

When the drive rise time is tightly constrained, it is important to follow
the ideal drive history as closely as possible.
Another way of presenting this calculation is as the scaled pressurization
rate $\tilde p$ (Fig.~\ref{fig:pptilde}).
To convert $\tilde p$ to an actual pressurization rate,
again choose the desired shock formation distance $l_0$ and divide 
$\tilde p$ by $l_0$ to find $\dot p$.
Thus, for a shock formation distance of at least 100\,$\mu$m,
the pressurization rate applied to Cu should be no more than about
3.5\,GPa/ns as the drive rises through, for instance, 10\,GPa
($\tilde p\simeq 350$\,GPa.$\mu$m/ns).
A convenient, but approximate, relation can be obtained between the 
drive pressure, sample thickness, and drive rise time
by dividing $\tilde p$ by $p$.
The resulting quantity has dimensions of speed:
distance to form a shock divided by fractional rate of change of pressure,
which is of similar order to the pulse length
(Fig.~\ref{fig:ptilde}).
A shock forms most quickly when this `shock formation speed' is lowest,
which is when the drive pressure is around the bulk modulus of the material.
For Cu, this speed is around 20\,km/s,
so a ramp of initial duration 10\,ns will form a shock after propagating through
of order 200\,$\mu$m of material.

For a given type of experiment, for example using a laser with a
limit on the pulse length,
the  most accurate measurement of evolution of ramp wave
usually require sample thicknesses to be significant
fractions of $l_0$.

\begin{figure}
\begin{center}\includegraphics[scale=0.72]{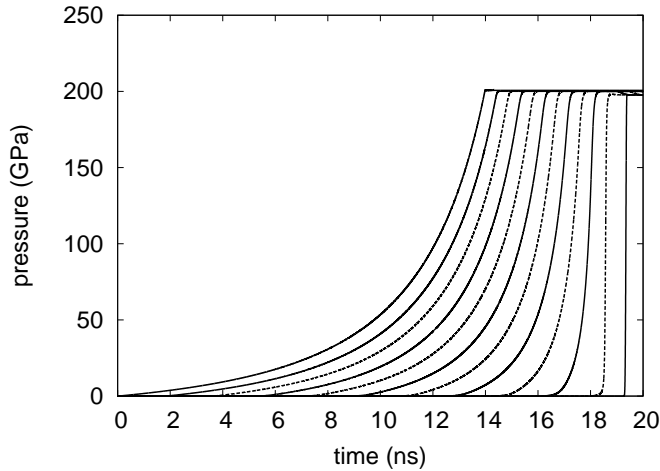}\end{center}
\caption{Pressure histories from spatially-resolved continuum dynamics
   simulation, showing shock formation in Al.
   The sample was driven using the ideal loading history, to 200\,GPa,
   with $l_0=100$\,$\mu$m.
   Pressure histories are shown for Lagrangian positions at intervals of 
   10\,$\mu$m from the loading surface.
   The shock formed simultaneously over the full pressure range at the
   100\,$\mu$m position, though with evidence of shock-smearing at the
   ends of the pressure range.
   The small pressure drop after shock formation is caused by the
   difference in compression between isentropic and shock compression.}
\label{fig:shockupsim}
\end{figure}

\begin{figure}
\begin{center}\includegraphics[scale=0.72]{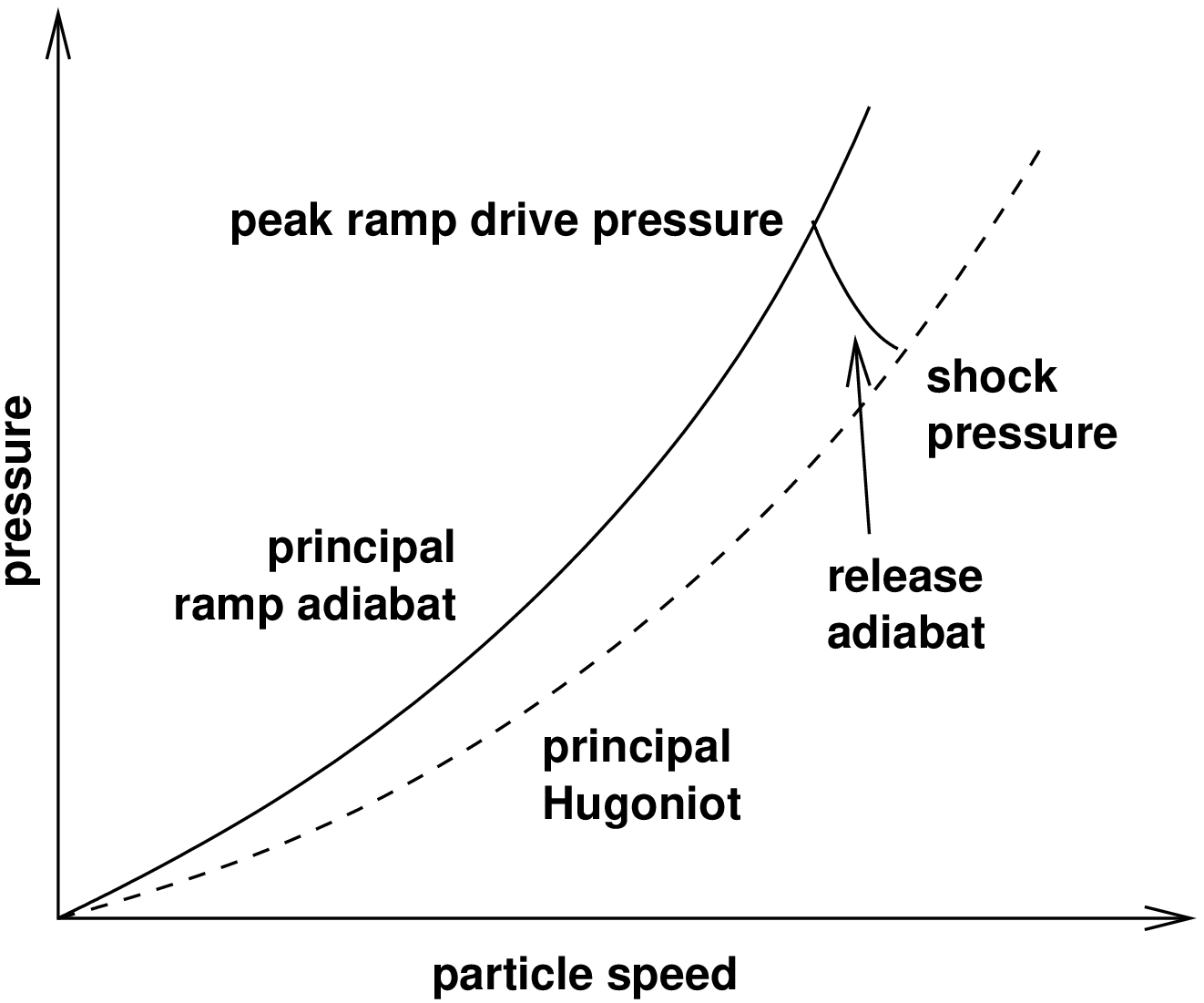}\end{center}
\caption{Schematic of shock formation in pressure-particle speed space,
   showing difference in particle speed that leads to partial unloading of
   the ramp-loaded material.
   For material loaded by the ramp wave, states move up the principal
   ramp adiabat to the peak ramp drive pressure.
   When the shock forms, a shock of the same pressure as the peak of the ramp
   would have a larger particle speed.
   The state in material loaded by a single shock lies on the
   principal Hugoniot.
   On formation of the shock, ramp-loaded and shock-loaded material
   must be at the same pressure and particle speed,
   so the ramp-loaded material re-expands down the release adiabat,
   and the shock pressure is slightly lower than the peak of the ramp.}
\label{fig:shockunload}
\end{figure}

Spatially-resolved continuum dynamics (hydrocode) simulations were performed 
of shock formation from a ramp drive in Cu,
using the ideal ramp shape calculated above.
The simulations used Lagrangian cells and a second order time-integration
algorithm of the predictor-corrector type.
Shocks were stabilized using artificial viscosity of the Wilkins and
von Neumann types: linear plus quadratic terms in the velocity gradient.
These numerical methods are well-established \cite{Benson92};
the computer program used was LAGC1D V6.0 \cite{LAGC1D}.
The spatial cells were 0.1\,$\mu$m wide.
Within the limitations of smearing from the artificial viscosity,
the ramp evolved into a shock simultaneously over the full pressure range,
and at the distance predicted by the characteristics analysis
(Fig.~\ref{fig:shockupsim}).
The shock pressure was slightly lower than than the top of the ramp
because the isentropic compression needed to reach a given pressure
is greater than the shock compression, so the ramp-loaded material
unloaded slightly into the shocked region (Fig.~\ref{fig:shockunload}).
This phenomenon is equivalent to the unloading produced when a high pressure
shock overtakes one of lower pressure \cite{Fritz95},
and has been discussed previously for shocks forming from a non-ideal ramp
\cite{Hayes04}.

\section{Propagation of an edge release across a ramp wave}
As with shock loading studies of material properties, ramp loading experiments
are usually intended to apply a one dimensional (1D) load to the sample 
over some useful, finite region.
The lateral extent of the 1D region is limited by the size of the driver
or the sample, e.g. the size of a laser spot, and 
also by lateral flow at the edges of the 1D region which propagate inward
as the ramp propagates through the sample.

\begin{figure}
\begin{center}\includegraphics[scale=0.72]{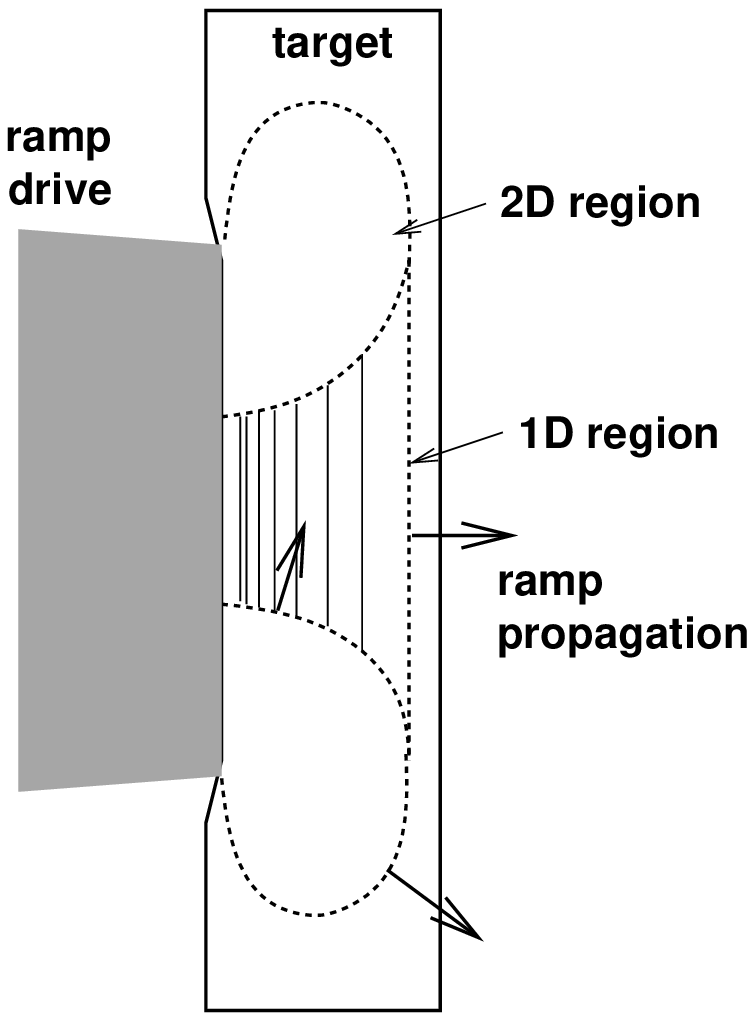}\end{center}
\caption{Schematic cross-section of ramp-loading experiment,
   showing lateral release propagating into ramp-loaded material.
   The edge release interacts with the continued application of the drive
   to produce a region of two-dimensional deformation that
   pinches off the one-dimensional ramp as time progresses.
   Solid contours in the one-dimensional region represent the
   ramp pressure, increasing with proximity to the drive.
   Thicker arrows show the direction of propagation of the waves.
   At any given time, the two-dimensional region has spread further laterally
   than axially because lateral propagation is through material of increasing
   sound speed.}
\label{fig:edgerel}
\end{figure}

Any infinitesimal increment of compression in the axial direction
propagates axially at the instantaneous longitudinal sound speed.
As this is the fastest mechanical signal supported by the material at that
compression, no signal from the edge can catch up with it.
However, laterally-propagating signals from the edge reduce the size of the 
1D region available for further axial increments of compression
(Fig.~\ref{fig:edgerel}).

The distance traveled laterally by signals traveling at state-dependent
speed $c$ through material compressed in the axial direction is
\begin{equation}
\Delta r = \int c\,dt.
\end{equation}
In general, this calculation is less useful than corresponding calculation
for a shock \cite{Swift-chablator-07}
because it is rare that a ramp would be ideal,
so integration has to be performed for the actual loading history
of a given experiment.
For the ideal loading history $p(\tau)$ implies a unique $c(\tau)$,
which allows $\Delta r(\tau)$ to be determined for a given material,
or $\Delta\tilde r(p)\equiv \Delta r/l_0$.
This calculation allows the aspect ratio of an experiment to be chosen,
to ensure that an adequate portion of the sample is subjected to
planar ramp loading.

Unlike the lateral release experienced by a shock, the release of an
ideally-shaped ramp is generally more gentle,
with an initially slow rate of release because of the initially slow rate
of pressurization.
The release may be further reduced when the load is generated by local
energy deposition (as in laser ablation) rather than inertial
confinement (such as a graded-density impactor \cite{Barker83}), 
because an elevated pressure is applied over the whole drive region.
The analysis presented above gives the region subjected to strictly
1D loading; in practice the initial perturbation will be small, and
experiments in which 2D release has started to take effect may not
be affected significantly.

\begin{figure}
\begin{center}\includegraphics[scale=0.72]{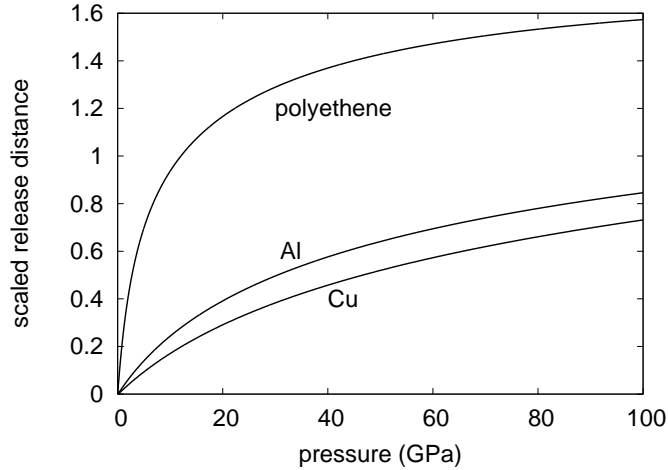}\end{center}
\caption{Scaled edge release at the drive surface, for ideal ramp loading.
   The scaled release distance is the lateral distance affected by the
   edge release, divided by the axial shock formation distance.}
\label{fig:edgepr}
\end{figure}

Calculations were made of the scaled release distance, again for
Al, Cu, and polyethene as prototype materials representative of
experiments on different metals and using plastic ablators
(Fig.~\ref{fig:edgepr}).
Thus for instance,
if Cu is loaded to 80\,GPa using the ideal ramp shape,
the scaled release distance is 0.6, meaning that the drive surface would
be affected by edge release within a distance of $0.6 l_0$ of the edge.
If the ramp rise time was chosen to give $l_0=100$\,$\mu$m
then the diameter of a laser drive spot should be at least 120\,$\mu$m
($2 l_0$, for release from opposite edges) plus the diameter of the
desired 1D region.

If the pressurization rate is slower than in the ideal ramp,
the region affected by edge release is larger.
The integration should however be done for the actual loading history used:
it is not generally accurate to scale by the overall rise time of the ramp
because $c$ generally varies nonlinearly with $p$.

\begin{figure}
\begin{center}\includegraphics[scale=0.66]{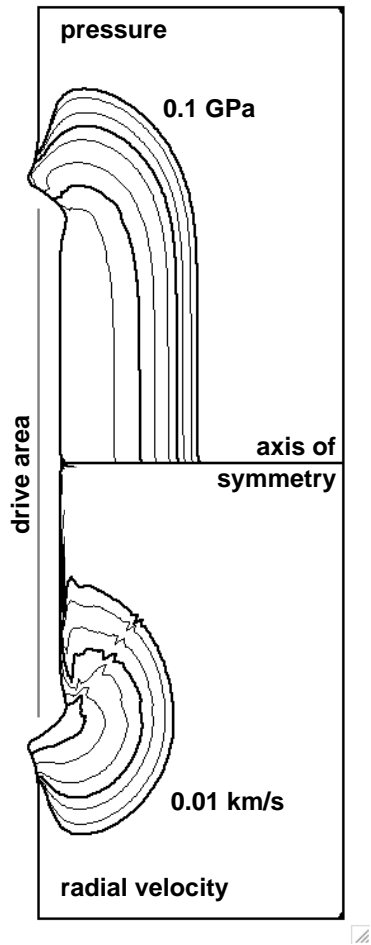}\end{center}
\caption{Spatially-resolved Lagrangian continuum dynamics simulation of
   edge release during ramp loading of Al.
   Loading was applied over a region 200\,$\mu$m in diameter at the left side
   (shown by the grey line), ramping linearly to zero over 5\,$\mu$m.
   The loading history was chosen to give a shock formation distance of
   100\,$\mu$m.
   The compression wave moves from left to right.
   The radius of the drive region was chosen to be equal to the
   shock formation distance, so that the scaled edge release distance
   is the fractional radius.
   This frame is at 9.8\,ns after the start of loading,
   when the drive pressure has reached 50\,GPa.
   Upper half shows pressure contours: 0.1, 0.2, 0.5, 1, 2, 5, 10, 20\,GPa
      increasing from right to left.
   Lower half shows contours of the lateral (outward) component of particle velocity:
      0.01, 0.02, 0.05, 0.1, 0.2, 0.5, 1\,km/s.
   Exact powers of ten are thicker lines.
   The furthest advanced point is at the same radius as the edge of the
   drive region.
   In the drive region, edge release is approximately 55\%\ of the
   way to the center, in agreement with the scaled edge release calculation.
   The simulations used a triangulated mesh with side lengths
   initially 2\,$\mu$m.}
\label{fig:edgerelsim}
\end{figure}

Two dimensional, spatially-resolved continuum dynamics (hydrocode) simulations 
were performed of ramp compression in Cu, using the ideal loading history.
Simulations were performed in two dimensions with Eulerian \cite{EUL2D} 
and Lagrangian \cite{LAGC} hydrocodes.
In both cases, the forward-time integration of the continuum equations
was finite difference over a staggered mesh
(particle velocity at nodes; material state at cell centers) 
using a second-order predictor-corrector numerical scheme;
the Eulerian simulations used third-order advection with the van Leer
flux limiter \cite{Benson92}.
Colormaps or contours of pressure showed reasonable agreement with the
characteristic analysis, but it was difficult to identify the onset
of release given the finite resolution of the continuum, and
pixellation and spatial averaging of the pressure field introduced
when generating graphics.
The progression of the edge release was clear in the radial
velocity component, demonstrating that the low-pressure compression was not
significantly eroded laterally, and that the region affected by lateral release 
was matched the characteristic analysis.
The radial velocity is a more direct measure of lateral release than is
the pressure, and was not affected by averaging in the contouring
algorithm.
(Fig.~\ref{fig:edgerelsim}.)

\section{Example calculations for laser-driven material dynamics experiments}
The applications and limitations of ramp loading using laser ablation
depend on the type of laser used.
We consider the following classes of laser:
\begin{description}
\item[High energy.]
Laser systems delivering $o(100)$\,J in a pulse.
Currently-operating examples include
TRIDENT at Los Alamos National Laboratory,
JANUS at Lawrence Livermore National Laboratory,
and OMEGA at the University of Rochester.
These are relatively large, building-sized facilities,
where experiments are performed at the facility.
Previous experimental work has established these systems as recognized
platforms for material dynamics studies, to a varying degree.
\item[Medium energy.]
Laser systems delivering $o(1)$\,J,in a pulse.
There are many such systems in existence.
They are usually much smaller, fitting within a small room.
It is often relatively straightforward to disassemble, crate, and re-assemble 
them, so it is feasible to transport them to other, fixed facilities
which may provide a particular range of in-situ measurements such as
diffraction from a synchrotron
\cite{McNaney08}.
However, the energy and pulse shaping is generally less suited to
material dynamics experiments.
\end{description}

In the following sections we apply the ramp loading analysis
to typical experimental configurations using these different classes of laser.

\subsection{High energy lasers in nanosecond shock mode}
Using JANUS and TRIDENT to generate shocks using ablation of nanosecond-scale
pulses \cite{Swift_elements_04},
the minimum rise time of the laser pulse is around 0.1\,ns.
Pressures of interest in material dynamics studies are typically
$\sim$10-100\,GPa.

For a metal sample, loaded by direct ablation of the sample itself, 
the scaled rise time $\tau=0.05-0.15$\,ns/$\mu$m, so
0.7-2\,$\mu$m of the sample is subjected to a ramp before a shock forms.
Samples are typically 10-200\,$\mu$m thick, so they are largely shocked.
A possible concern is diffraction from driven side if the x-ray penetration
depth is not much greater than the shock formation distance.

If the sample is loaded by ablation of a plastic ablator, such as
parylene \cite{Swift-chablator-07},
$\tau=0.2-0.3$\,ns/$\mu$m,
so 0.3-0.5\,$\mu$m of the ablator is subjected to a ramp.
This thickness is small compared with typical ablator thicknesses 
of 10-20\,$\mu$m.

Edge release is not relevant during the shock formation stage: 
it affects a tiny region compared with typical laser spot sizes of
1-10\,mm diameter.

\subsection{High energy lasers in nanosecond ramp mode}
The TRIDENT laser has been used previously to induce ramp loading
with a shaped pulse up to 2.5\,ns long \cite{Swift_lice_05}.
For metal samples and a peak pressure of a few tens of gigapascals,
a shock would form in 15-50\,$\mu$m using the ideal loading history.
The radial extent of the region affected by edge release would be 0.1-0.8
of this, which is small compared with typical TRIDENT drive spot diameters 
of 1-5\,mm.

More recently, TRIDENT and JANUS have been modified to allow shaped pulses 
10-20\,ns long.
For otherwise similar experiments, a shock would form in 60-400\,$\mu$m.
Care may be needed to control the extent of edge release when operating 
with a drive spot of diameter 1\,mm or less.

\subsection{High energy lasers in microsecond mode}
The TRIDENT laser system can be operated in a frustrated amplification mode
in which the pulse may be varied from around 50\,ns to many microseconds.
These long pulses may be used to ablate material confined by
a transparent tamper \cite{Colvin07,Paisley08,Loomis07}.
Pressures have been limited by breakdown of the tamper,
and could likely be extended to higher pressures or longer durations by
better spatial smoothing of the laser beam.
Pressures of 10\,GPa have been demonstrated, sustained for hundreds of
nanoseconds.
The pulse shape can be varied to induce shocks and ramps, among other 
shapes.

For shock loading, a Pockels cell has been used to clip the early part of
the pulse, producing a minimum rise time of a few nanoseconds.
The initial loading history is therefore a ramp, forming a shock
in a thickness of around 100\,$\mu$m.

Ramp loading has been demonstrated using a Gaussian pulse history of 
160\,ns full width, half maximum.
Using the ideal loading history, a shock would form in $\sim$3\,mm.

The edge release distance is around 0.2 of the shock formation distance,
which is manageable for typical drive spot diameters of 5-8\,mm.

Long pulses at TRIDENT have also been used to accelerate laser flyers
for impact experiments \cite{Swift-flyer-05}.
To minimize heating and damage in the flyer,
the compression wave should preferably not induce a shock,
so the shock formation distance should be greater than the flyer thickness.
During the acceleration process, the ablation pressure typically
reaches a maximum of around 0.1\,GPa.
Flyers have typically been 0.1-1\,mm thick.
The scaled rise time $\tau=5\times 10^{-4}$\,ns/$\mu$m,
implying a minimum rise time of 0.05-0.5\,ns, which is 
far shorter than those generally used.
The effects of edge release should be negligible for these pressures, 
so edge release of the ramp through the flyer thickness 
should not contribute significantly to curvature in the flyer.

\subsection{Medium energy lasers}
A portable loading laser has been used at the Advanced Photon Source synchrotron
at Argonne National Laboratory to induce dynamic loading in samples
with in-situ probing by synchrotron x-rays \cite{McNaney08}.
The laser pulse had a Gaussian temporal profile of
12\,ns full-width, half-maximum.
The focal spot used to load the sample had a diameter of 250-300\,$\mu$m.
The laser pulse energy was around 0.4\,J.
Using a plastic ablator, the pressure induced in an Al sample
should be in the range 1-10\,GPa,
implying a scaled rise time of $\tau=0.05-0.2$\,ns/$\mu$m for shock formation.
Thus a shock would form beyond a plastic thickness of 60-240\,$\mu$m,
which is thicker than typical for the ablator ($\sim$20\,$\mu$m).
In the sample itself, with or without an ablator,
the scaled rise time is $\tau=0.01-0.05$\,ns/$\mu$m,
so a shock would form beyond a thickness of 240-1200\,$\mu$m.
This too is greater than the samples used, and thicker than a shock could
be supported by that laser pulse length,
so the drive was a ramp for all practical purposes.

The edge release region in typical ablators was much smaller 
than the drive spot.
The edge release in typical samples ($\sim$100\,$\mu$m thick) was also small.

\section{Conclusions}
We have derived expressions for ramp loading in compact, scaled form,
allowing the adiabat for any material to be used to predict the distance
for any arbitrary ramp to steepen into a shock.
The calculation can be performed for adiabats expressed in tabular form,
derived from material models of arbitrary complexity.
The steepening relation can be used to determine the `ideal' scaled loading
history for a material, maximizing the distance for the shock to form.
Steepening relations were derived for Al, Cu, and polyethene,
as material representative of types commonly used in material dynamics studies.
The propagation of lateral release waves across a ramp was analyzed,
giving scaled relations for the region affected by lateral release when
axial loading follows the ideal history.
The analyses proceeded by considering characteristics;
hydrocode simulations were used to verify the accuracy of the analyses.

The ramp evolution and edge release analyses were applied to
situations in several types of laser loading experiment.
It was demonstrated that properly-designed shock experiments 
at large-scale laser facilities do not subject unduly large amounts
of the sample to ramp rather than shock loading from the finite rise time
of the laser pulse, which has previously been a concern.
The use of plastic ablators in particular eliminates any ramp region
from the sample.
The calculations also capture the limitations of ramp loading and the extent
of lateral release in these experiments in a compact form, without
requiring spatially-resolved simulations in two or even one dimension.

\section{Acknowledgments}
This work was performed in support of
Laboratory-Directed Research and Development projects 06-SI-004
(Principal Investigator: Hector Lorenzana)
and 08-ER-038 (Principal Investigators: Damian Swift and Bassem El-Dasher),
under the auspices of
the U.S. Department of Energy under contract
DE-AC52-07NA27344.

\end{document}